\documentclass[twocolumn,aps,prb,superscriptaddress,floatfix]{revtex4}
\usepackage{graphicx}
\usepackage{dcolumn}
\usepackage{bm}
\usepackage{xfrac}
\begin{document}
%

\title{Magnetic Excitations in Square Lattice Iridates:\\ Contrast between Ba$_2$IrO$_4$ and Sr$_2$IrO$_4$}

\author{J. P. Clancy}
\affiliation{Department of Physics, University of Toronto, Toronto, Ontario, M5S 1A7, Canada}

\author{H. Gretarsson}
\affiliation{Department of Physics, University of Toronto, Toronto, Ontario, M5S 1A7, Canada}

\author{A. Lupascu}
\affiliation{Department of Physics, University of Toronto, Toronto, Ontario, M5S 1A7, Canada}

\author{J. A. Sears}
\affiliation{Department of Physics, University of Toronto, Toronto, Ontario, M5S 1A7, Canada}

\author{Z. Nie}
\affiliation{Department of Physics, University of Toronto, Toronto, Ontario, M5S 1A7, Canada}

\author{M. H. Upton}
\affiliation{Advanced Photon Source, Argonne National Laboratory, Argonne, Illinois 60439, USA}

\author{Jungho Kim}
\affiliation{Advanced Photon Source, Argonne National Laboratory, Argonne, Illinois 60439, USA}

\author{Z. Islam}
\affiliation{Advanced Photon Source, Argonne National Laboratory, Argonne, Illinois 60439, USA}

\author{M. Uchida}
\affiliation{Laboratory of Atomic and Solid State Physics, Department of Physics, Cornell University, Ithaca, New York 14853, USA}

\author{D. G. Schlom}
\affiliation{Department of Materials Science and Engineering, Cornell University, Ithaca, New York 14853, USA}
\affiliation{Kavli Institute at Cornell for Nanoscale Science, Ithaca, New York 14853, USA}

\author{K. M. Shen}
\affiliation{Laboratory of Atomic and Solid State Physics, Department of Physics, Cornell University, Ithaca, New York 14853, USA}
\affiliation{Kavli Institute at Cornell for Nanoscale Science, Ithaca, New York 14853, USA}

\author{Young-June Kim}
\affiliation{Department of Physics, University of Toronto, Toronto, Ontario, M5S 1A7, Canada}

\begin{abstract}
We report a resonant inelastic x-ray scattering (RIXS) investigation of ultra-thin epitaxial films of Ba$_2$IrO$_4$, and compare their low energy magnetic and spin-orbit excitations to those of their sister compound Sr$_2$IrO$_4$. Due to the 180$^\circ$ Ir-O-Ir bond, the bandwidth of the magnon and spin-orbiton is significantly larger in Ba$_2$IrO$_4$, making it difficult to describe these two types of excitations as separate well-defined quasiparticles. Both types of excitations are found to be quite sensitive to the effect of epitaxial strain. In addition, we find that the d-level inversion observed in Sr$_2$IrO$_4$ is absent in Ba$_2$IrO$_4$, as predicted in recent theoretical studies. Our results illustrate that the magnetic properties of Ba$_2$IrO$_4$ are substantially different from those of Sr$_2$IrO$_4$, suggesting that these materials need to be examined more carefully with electron itinerancy taken into account.
\end{abstract}


\maketitle

\section{introduction}

Iridium-based transition metal oxides have emerged as a rich source for novel phenomena driven by strong spin-orbit coupling (SOC) effects\cite{BJKim2008,BJKim2009,Jackeli2009,Cao2018,Bertinshaw2019}.  In particular, the layered perovskite iridate Sr$_2$IrO$_4$ is considered as the prototypical $j_{eff}$ = $\sfrac{1}{2}$ spin-orbit Mott insulator, whose ground state arises due to a complex interplay between crystal electric field, SOC, and electronic correlation \cite{BJKim2008}. In addition, the strong parallels between Sr$_2$IrO$_4$ and La$_2$CuO$_4$, the parent compound of the high T$_c$ cuprates, have led to natural associations with superconductivity\cite{Wang2011,Watanabe2013,JKim2012a,Yang2014} and have inspired extensive research into the properties of doped Sr$_2$IrO$_4$.  However, in spite of several theoretical predictions, the experimental realization of iridate superconductivity through chemical substitution has proven elusive \cite{Wang2011,Watanabe2013,Yang2014}. Physical properties and key issues of Sr$_2$IrO$_4$ have been discussed in recent review articles \cite{Cao2018,Bertinshaw2019}.

Despite great interest in Sr$_2$IrO$_4$, its sister compound Ba$_2$IrO$_4$ has received comparatively little experimental attention, largely due to the difficulty of sample synthesis. Since Ba$_2$IrO$_4$ is metastable under ambient conditions, high pressure (P $>$ 6 GPa) synthesis is required to obtain bulk Ba$_2$IrO$_4$ samples \cite{Okabe2011}. A quick glance at the existing data seems to suggest that these two compounds are magnetically similar. Ba$_2$IrO$_4$ and Sr$_2$IrO$_4$ display identical magnetic transition temperatures (T$_N$ = 240 K) and very similar magnetically ordered structures \cite{Boseggia2013}. Angle-resolved photoemission (ARPES) and x-ray absorption spectroscopy studies report that the electronic structures of these two compounds are also quite similar \cite{Moser2014,MorettiSala2014b,Uchida2014}. This is somewhat surprising given that there exists an important structural difference between the two materials. The IrO$_6$ octahedra are rigidly aligned in Ba$_2$IrO$_4$, just like the CuO$_6$ octahedra in cuprates, while these octahedra are rotated by $\sim$ 11$^{\circ}$ about the {\it c}-axis in Sr$_2$IrO$_4$. This gives rise to key structural differences in terms of Ir-O-Ir bond angles (180$^{\circ}$ vs. 158$^{\circ}$) and local distortions of the IrO$_6$ octahedra (7\% vs. 4.5\% axial elongation). The straight bond is expected to increase hybridization and make Ba$_2$IrO$_4$ more itinerant, which puts into question the applicability of the $j_{eff}$=1/2 pseudo-spin model for this material. Indeed, recent theoretical studies have pointed out the shortcomings of the pseudo-spin model \cite{Solovyev2015} and significant mixing of the $t_{2g}$, $e_g$, and oxygen $p$ states \cite{Rosciszewski2016}. Increased hybridization also suggests a stronger superexchange interaction between the pseudo-spins in Ba$_2$IrO$_4$, which seems to be supported by theoretical calculations \cite{Katukuri2012,Solovyev2015}. A recent two-magnon Raman scattering study also seems to indicate the superexchange in Ba$_2$IrO$_4$ is about 20 \% to 30 \% larger than that in Sr$_2$IrO$_4$.\cite{Tsuda2016} Remarkably, this means that the increase in the superexchange is {\em exactly} compensated by the anisotropic terms in the magnetic Hamiltonian to keep $T_N$ unchanged.

The subtlety of the magnetic anisotropy due to the local distortion was discussed in the theoretical study by Bogdanov and coworkers \cite{Bogdanov2015}. On a local level, competition between octahedral distortions and longer-range ligand-field effects from the highly charged Ir$^{4+}$ ions is believed to reverse the order of the low-lying d-levels in Sr$_2$IrO$_4$, but not in Ba$_2$IrO$_4$.
The net result is to suppress the effect of the tetragonal distortion in Ba$_2$IrO$_4$  \cite{Bogdanov2015}. However, the weak Ising exchange anisotropy could not explain the basal plane ordering direction observed experimentally \cite{Boseggia2013}, and various additional interactions such as the interlayer exchange interaction \cite{Katukuri2014c}, single-ion anisotropy, and pseudo-quadrupolar interaction \cite{Hou2016} have been suggested to explain the magnetic structure. In addition, Solovyev et al. suggested that the pseudo-spin model is inadequate for Ba$_2$IrO$_4$ due to the increased hybridization, and higher-order terms arising from itinerancy of the system should be considered \cite{Solovyev2015}. Unfortunately, there have been only limited experimental studies on Ba$_2$IrO$_4$ to test these theories.

In this paper, we examine the magnetic Hamiltonian of Ba$_2$IrO$_4$ by taking advantage of two technical advances made in recent years. Epitaxial film growth \cite{Nichols2014,Uchida2014,Zhao2021} offers an elegant alternative to the high pressure synthesis techniques required for Ba$_2$IrO$_4$\cite{Okabe2011}, providing a method for both stabilizing the crystal structure and exerting a high level of control over key structural parameters. However, probing the characteristic excitations of such thin film samples presents a significant experimental challenge. We use resonant inelastic x-ray scattering (RIXS) to map out the dispersion of magnetic and orbital excitations in ultrathin films of Ba$_2$IrO$_4$ ($\sim$13 to 20 nm). RIXS is a momentum-resolved and element-specific probe of elementary excitations such as magnons or orbitons \cite{Braicovich2009,Schlappa2012,JKim2012a}. We found that despite the overall similarity of the magnetic excitation spectra, the magnetic energy scales in Ba$_2$IrO$_4$ differ significantly from those in Sr$_2$IrO$_4$. The magnon and spin-orbital excitations are no longer well separated in Ba$_2$IrO$_4$, making it difficult to justify the use of the pseudo-spin model to describe magnetic properties. In addition, we confirm that the magnetic anisotropy is reversed in Ba$_2$IrO$_4$ as predicted in Ref.~\onlinecite{Bogdanov2015}. Our results show that the structural differences between square lattice iridates make the magnetic properties of Ba$_2$IrO$_4$ distinct from those of Sr$_2$IrO$_4$.


\section{Experimental Details}

\begin{figure}
\includegraphics[width=3.1in]{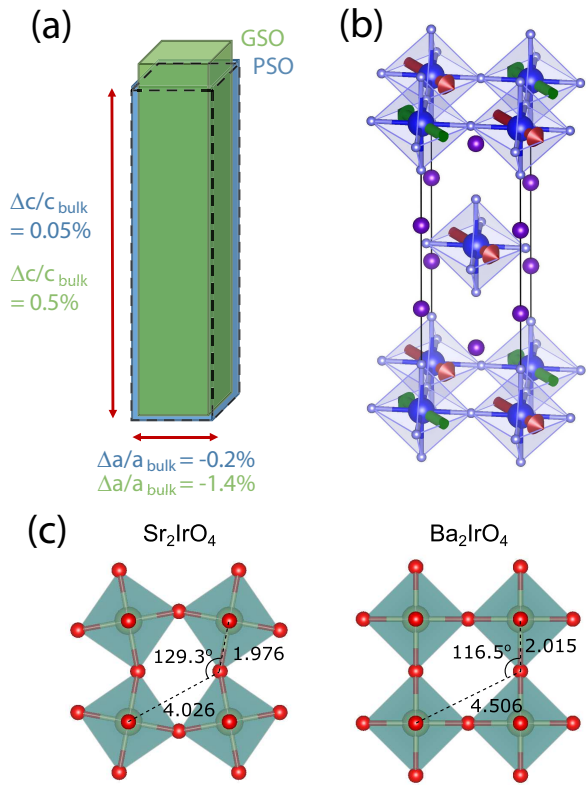}
\caption{(a) Effect of epitaxial strain on the crystal structure of Ba$_2$IrO$_4$. The unit cell of bulk Ba$_2$IrO$_4$ is illustrated by the dashed black lines. Epitaxial strains generated by PrScO$_3$ (PSO, blue) and GdScO$_3$ (GSO, green) substrates have been magnified by a factor of 10. (b) The collinear antiferromagnetic ground state of Ba$_2$IrO$_4$. (c) A comparison of Ir-O-Ir bond geometry in Sr$_2$IrO$_4$ and Ba$_2$IrO$_4$}
\label{fig_structure}
\end{figure}

Measurements were performed on two ultrathin Ba$_2$IrO$_4$ samples prepared by molecular beam epitaxy: one 20 nm ($\sim$ 15 unit cell) film grown on a PrScO$_3$ (PSO) substrate, and one 13 nm ($\sim$ 10 unit cell) film grown on GdScO$_3$ (GSO).
A more detailed description of the synthesis procedure can be found in Reference \onlinecite{Uchida2014}.  Reflection high-energy electron diffraction (RHEED) was used to monitor the films throughout the growth process, and to verify that there was no evidence of lattice relaxation.  After synthesis, a combination of low-energy electron diffraction (LEED) and x-ray diffraction was used to confirm the phase purity, crystallinity, and lattice constants of the films. In addition, the ARPES data \cite{Uchida2014} provides confirmation that the electronic structure of thin film samples are comparable to bulk crystals. The degree of lattice mismatch in these two samples is dramatically different, with the PSO sample approaching unstrained bulk material ($\Delta a/a$ = -0.2\%) and the GSO sample exhibiting a sizeable compressive strain ($\Delta a/a$ = -1.4\%).  The effect of epitaxial strain on the crystal structure of Ba$_2$IrO$_4$ is illustrated in Figure~\ref{fig_structure}(a). Bulk single crystals of Sr$_2$IrO$_4$  were synthesized using the self-flux technique, as described in Reference~\cite{Cao1998}.

Resonant elastic x-ray scattering (REXS) measurements at the Ir L$_3$-edge were performed using Beamline 6-ID-B at the Advanced Photon Source.  Measurements were carried out in vertical scattering geometry, with the polarization of the incident beam perpendicular to the plane defined by {\bf k$_i$} and {\bf k$_f$} (i.e. $\sigma$-polarization).  Polarization analysis of the scattered beam was performed using the (0,0,8) reflection from a pyrolytic graphite analyzer crystal.

Ir L$_3$-edge RIXS measurements were performed using the MERIX spectrometer on Beamline 30-ID-B at the Advanced Photon Source.  A channel-cut Si-(8,4,4) secondary monochromator, and spherical (2 m radius) diced Si-(8,4,4) analyzer crystal were used to obtain an overall energy resolution of 45 meV (FWHM). To reduce the strong elastic scattering contribution from sample and substrate, measurements were carried out in horizontal scattering geometry, with a scattering angle close to 2$\theta$ = 90$^{\circ}$ and at $T=10$~K.  To maximize the inelastic scattering from the films, measurements were performed near grazing incidence, with an angle of incidence $\alpha$ $<$ 1$^{\circ}$.

\section{Experimental Results and Discussion}

\subsection{Magnetic Structure}

\begin{figure}
\includegraphics[width=3.1in]{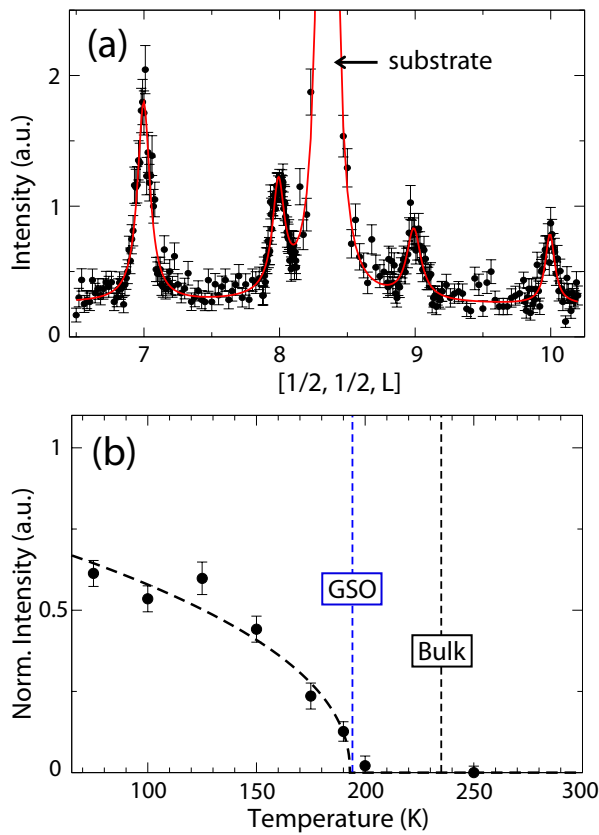}
\caption{(Color online) Resonant elastic x-ray scattering from 13 nm thin film Ba$_2$IrO$_4$ grown on a GSO substrate. (a) A scan showing magnetic Bragg peaks along the [1/2, 1/2, L] direction. These measurements were collected in the $\sigma$-$\pi$' polarization channel at T = 7 K. (b) Temperature dependence of the integrated intensity of the (1/2, 1/2, 7) magnetic Bragg peak. }
\label{fig_bragg}
\end{figure}

The magnetic structure of both samples was examined using REXS. Figure~\ref{fig_bragg} shows Ir L$_3$-edge REXS data on our most highly strained film, Ba$_2$IrO$_4$ on GSO. As shown in Fig.~\ref{fig_bragg}(a), magnetic Bragg peaks were observed at [1/2, 1/2, L] wave vectors for all integer values of L. These measurements were collected well within the magnetically ordered phase (T = 7 K), using polarization analysis to distinguish magnetic scattering ($\sigma$-$\pi$') from charge scattering ($\sigma$-$\sigma$'). This magnetic selection rule is consistent with the magnetic structure of bulk Ba$_2$IrO$_4$ as reported by Boseggia {\it et al.}\cite{Boseggia2013}: a collinear {\it ab}-plane antiferromagnet.

The observation of magnetic peaks at both L = {\it odd} and L = {\it even} implies the presence of multiple magnetic domains, with ordered moments aligning along either of the crystallographically equivalent [110] or [$\overline{1}$10] directions.  The same magnetic selection rule is also observed for Ba$_2$IrO$_4$ on PSO, which suggests that, as in the case of thin film Sr$_2$IrO$_4$\cite{Lupascu2014}, epitaxial strain does not perturb the magnetic structure of the bulk material. As in previous measurements on thin film Sr$_2$IrO$_4$\cite{Lupascu2014}, we note that the magnetic correlations in thin film Ba$_2$IrO$_4$ are highly anisotropic.  Observed magnetic correlation lengths are ~100 times longer within the ab-plane than along the c-axis stacking direction.

In Fig.~\ref{fig_bragg}(b), the temperature dependence of the (1/2, 1/2, 7) magnetic Bragg peak is shown. A limited number of temperature points were obtained, which prevented us from detailed analysis of the phase transition. However, it is clear that the magnetic transition temperature in this compressively strained is suppressed compared to that in a bulk sample. This observation is consistent with the behavior observed for Sr$_2$IrO$_4$\cite{Lupascu2014}, in which suppression of $T_N$ was observed for the films prepared with compressive epitaxial strain.

\begin{figure*}
\includegraphics[width=6.2in]{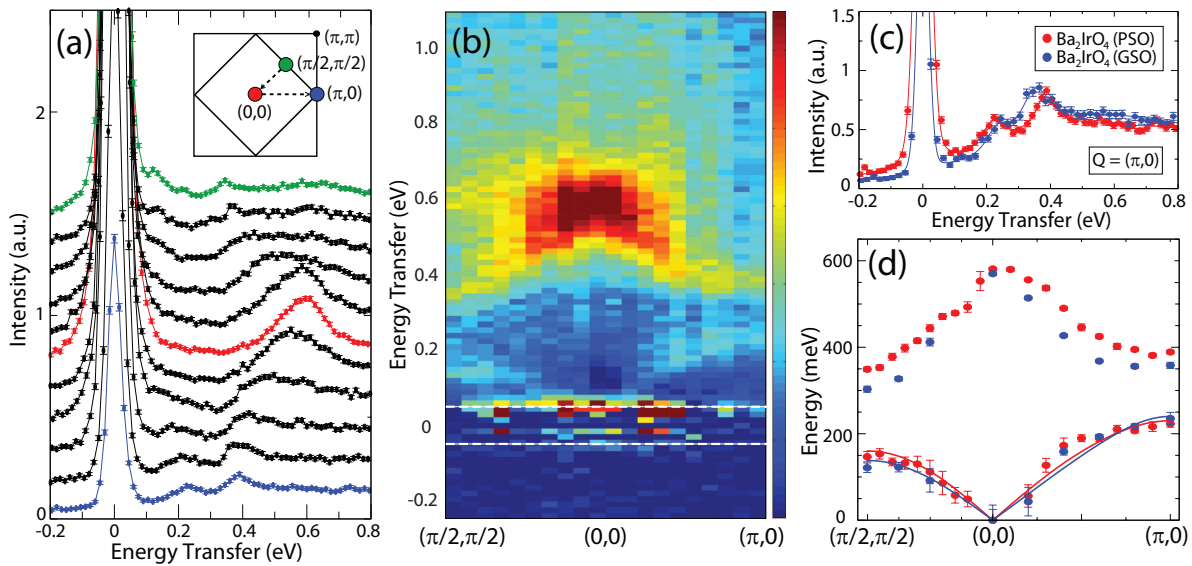}
\caption{(a) Representative energy scans performed on 20 nm thin film Ba$_2$IrO$_4$ grown on a PSO substrate. Colored spectra were obtained at the high symmetry points of the magnetic Brillouin zone shown in the inset. The top half of the spectra corresponds to the scans obtained along the $(0, 0)-(\pi/2, \pi/2)$ momentum transfer direction, while the bottom half were obtained for $(0, 0)-(\pi, 0)$. Spectra have been offset vertically for illustrative purposes. Inset: The magnetic Brillouin zone for Ba$_2$IrO$_4$. (b) Colour contour map showing the dispersion of the low energy magnon ($\sim$0 to 250 meV) and spin-orbit exciton ($\sim$350 to 650 meV) modes in Ba$_2$IrO$_4$. The elastic scattering contribution has been subtracted from this panel in order to highlight the inelastic features. The position of the elastic line is denoted by dashed white lines. (c) Strain dependence of magnetic and orbital excitations in Ba$_2$IrO$_4$ grown on PSO (-0.2\% strain, red) and GSO (-1.4\% strain, blue) substrates. (d) Dispersion of magnon and spin-orbit exciton modes in thin film Ba$_2$IrO$_4$. The magnetic dispersion can be fit to a Heisenberg $J-J'-J''$ model (solid red and blue lines).}
\label{fig_rixs}
\end{figure*}

\subsection{Excitations}
Results from the RIXS measurements on the essentially unstrained film, Ba$_2$IrO$_4$ on PSO, are presented in Fig.~\ref{fig_rixs}.  Figure~\ref{fig_rixs}(a) shows representative energy scans collected at high symmetry points throughout the magnetic Brillouin zone.  Here the scattering intensity is plotted as a function of energy transfer, $\Delta$E.  The distinguishing features of these spectra are: (1) a strong, resolution-limited elastic line ($\Delta$E = 0), (2) a dispersive magnetic excitation, or magnon, at low energy ($\Delta$E $<$ 250 meV), and (3) a dispersive orbital excitation, or spin-orbit exciton, at higher energy (350 $<$ $\Delta$E $<$ 650 meV).  Figure~\ref{fig_rixs}(b) shows a color intensity map constructed out of energy scans similar to those of Fig.~\ref{fig_rixs}(a).  In order to emphasize the inelastic features, the elastic scattering contribution has been subtracted from the data sets.

The strain dependence of the excitation spectrum of Ba$_2$IrO$_4$ is illustrated in Fig.~\ref{fig_rixs}(c), which compares representative energy scans performed on Ba$_2$IrO$_4$ films.  These scans have been collected at the ($\pi$,0) zone boundary position, where the energy of the magnon mode is at its maximum.  Note that epitaxial strain has a significant impact on both the magnetic and orbital excitations of this material.  As the magnitude of the compressive strain grows larger, the energy of the magnon mode is enhanced, while that of the spin-orbit exciton mode is reduced.  Similar strain-induced tuning of the zone boundary magnon energy has also been observed in thin film Sr$_2$IrO$_4$\cite{Lupascu2014}, where the magnetic bandwidth can be lowered (raised) by the application of a moderate tensile (compressive) strain.  However, by measuring the full dispersion of the magnetic excitations, as shown in Fig.~\ref{fig_rixs}(d), a more detailed picture of the strain dependence emerges. Each scan at a given momentum was fit using a fit function consisting of elastic-line (resolution-limited pseudo-Voight), magnon (Gaussian), spin-orbit exciton (Lorentzian, Gaussian), and electron-hole continuum/background (broad Gaussian). Only magnon and spin-orbit exciton positions are plotted in Fig.~\ref{fig_rixs}(d).

\begin{figure}
\includegraphics[width=3in]{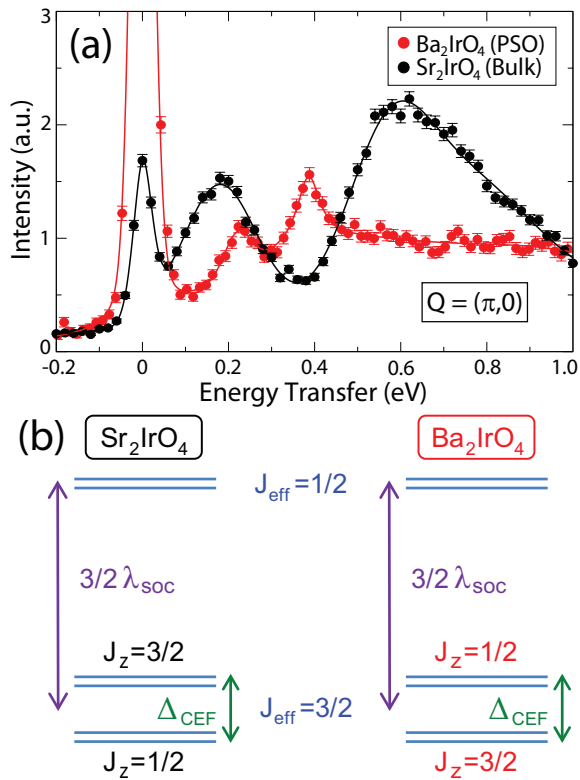}
\caption{ (a) Representative excitation spectra measured at the ($\pi$,0) zone boundary position for thin film Ba$_2$IrO$_4$ (20 nm, PSO substrate) and bulk single crystal Sr$_2$IrO$_4$.  Both spectra were obtained using grazing incidence geometry. (b) The orbital energy level schemes implied by the spectra in (a). Note that the sign of the non-cubic crystal field splitting must be opposite in Ba$_2$IrO$_4$ ($\Delta_{CEF}$ $>$ 0) and Sr$_2$IrO$_4$ ($\Delta_{CEF}$ $<$ 0). }
\label{fig_comp}
\end{figure}

A direct comparison of Ba$_2$IrO$_4$ and Sr$_2$IrO$_4$ spectra, obtained under equivalent experimental conditions, is provided in Fig.~\ref{fig_comp}(a).  The energy of the magnon mode at {\bf Q} = ($\pi$,0), which provides a measure of the full magnetic bandwidth, is significantly larger in Ba$_2$IrO$_4$ (223$\pm$4 meV) than Sr$_2$IrO$_4$ (178$\pm$4 meV).  This difference in bandwidth ($\sim$25\%) is qualitatively consistent with predictions based on quantum chemistry calculations by Katukuri {\it et al}\cite{Katukuri2012}. One important consequence of the larger bandwidth in Ba$_2$IrO$_4$ is the proximity of the
magnon mode and the spin-orbiton mode. These two modes are well separated in Sr$_2$IrO$_4$, but the larger bandwidth in Ba$_2$IrO$_4$ pushes the magnon mode up and the spin-orbiton mode down as shown in Fig.~3(a). This is also clearly shown in the dispersion plot of Fig.~2(d), in which these two modes seem to come close to each other in energy as they approach the $(\pi,0)$ point. In many other iridates described by the $j_{eff}=1/2$ picture, the  $j_{eff}=1/2$ and $j_{eff}=3/2$ states are well separated, meaning that the magnon and spin-orbiton can be treated as separate quasi-particles. In Ba$_2$IrO$_4$, however, lack of such a clear separation between the $j_{eff}=1/2$ and $j_{eff}=3/2$ states indicate that full understanding of these collective excitations will require consideration of all 6 bands in t$_{2g}$ manifolds.

With the above caveat in mind, we will discuss these excitations separately here in order to provide comparison with existing literature. To analyze the magnetic dispersion relation, we employ a $J-J'-J''$ pseudo-spin model.  This is the same phenomenological model which has previously been used to describe magnon dispersion in Sr$_2$IrO$_4$ \cite{JKim2012a} ($J$ = 60, $J'$ = -20, $J''$ = 15 meV). Substantial ferromagnetic next-nearest-coupling was introduced to account for the observed large zone-boundary dispersion. Similarly sizable zone-boundary dispersion is also observed for Ba$_2$IrO$_4$, and the dispersion relation is reasonably well described with $J \approx 85$, $J' \approx -15$, $J'' \approx 10$~meV for the PSO film. Similar parameters are also obtained from fitting the GSO film data ($J \approx 82$, $J' \approx -19$, $J'' \approx 16$~meV). The much larger value of $J$ in Ba$_2$IrO$_4$ can be understood from the 180$^\circ$ Ir-O-Ir bond angle according to the Goodenough-Kanamori rule. However, the further neighbor interactions in Ba$_2$IrO$_4$ are almost as large as the values in Sr$_2$IrO$_4$. This is difficult to understand in the superexchange picture, since the next-nearest-neighbor hopping paths in Ba$_2$IrO$_4$ are very different from those in Sr$_2$IrO$_4$ due to the absence of IrO$_6$ octahedral rotations (See Fig.1(c)). It is possible that superexchange involving $j_{eff}=3/2$ might account for this, although no such a calculation has been reported. However, we would like to point out that the large zone-boundary dispersion could be explained using so-called ring exchange as discussed in the context of La$_2$CuO$_4$ \cite{Coldea2001}, which could indicate considerable electron itinerancy (breakdown of the assumption of large Hubbard $U$).
It should be noted that recent {\it ab initio} calculations have demonstrated that inter-plane and symmetric anisotropic interactions are required in order to reproduce the observed magnetic structure of Ba$_2$IrO$_4$\cite{Katukuri2014c}. However the small size of these terms (estimated to be $\sim$0.5 to 3.5 meV) means that their contribution to the dispersion is overshadowed by the much stronger in-plane isotropic exchange interactions.

These results demonstrate that rather than simply tuning the magnetic bandwidth, epitaxial strain has a much more subtle effect on the balance of magnetic interactions.  Although compressive strain does serve to increase the overall bandwidth, it actually does so by reducing the nearest-neighbor exchange interaction, while enhancing the second and third neighbor interactions.

There are also notable differences in the orbital excitations of these two materials.  The spin-orbit exciton mode, originally observed by Kim {\it et al.}\cite{JKim2012a}, corresponds to excitations between the $j_{eff}$ = $\sfrac{3}{2}$ and $j_{eff}$ = $\sfrac{1}{2}$ states, which acquire dispersion as they propagate against an antiferromagnetically ordered background.  Recent RIXS measurements on Sr$_2$IrO$_4$ have demonstrated that this mode exhibits a strong dependence on the photon polarization \cite{JKim2014}. When the sample is rotated with respect to the incoming photon polarization, the RIXS matrix elements will change as follows: Scattering from wavefunctions with a larger in-plane orbital component is enhanced (suppressed) in grazing (normal) incidence, while scattering from wavefunctions with a larger out-of-plane orbital component is suppressed (enhanced).  Applying this to the ideal $j_{eff}$ = 3/2 wavefunctions, which can be expressed as
$| \sfrac{3}{2}, \pm \sfrac{3}{2} \rangle$ = $(\mp |d_{yz}, \pm \sigma \rangle + i | d_{xz}, \pm \sigma \rangle)/ \sqrt{2} $ and
$| \sfrac{3}{2}, \pm \sfrac{1}{2} \rangle$ = $(\mp |d_{yz}, \mp \sigma \rangle + i | d_{xz}, \mp \sigma \rangle + 2 | d_{xy}, \pm \sigma \rangle)/ \sqrt{6} $, we expect grazing incidence to result in a preferential enhancement of excitations associated with the $j_z$ = $\pm$\sfrac{1}{2} states.

As illustrated in Fig.~\ref{fig_comp}(a), the orbital excitations of Ba$_2$IrO$_4$ in grazing incidence are markedly different from those of Sr$_2$IrO$_4$ measured in the same geometry, but bear a striking resemblance to those of Sr$_2$IrO$_4$ measured in normal incidence (See Fig.~2b in Ref.~\onlinecite{JKim2014}).  The grazing incidence data in Fig.~\ref{fig_comp}(a) implies that the $j_z$ = $\pm$\sfrac{1}{2} levels form the lower energy $j_{eff}$ = $\sfrac{3}{2}$ states in Ba$_2$IrO$_4$, but the higher energy states in Sr$_2$IrO$_4$.  This energy level scheme is illustrated in Fig.~\ref{fig_comp}(b).  Note that this scheme implies that the sign of the effective non-cubic crystal field splitting ($\Delta_{CEF}$) must be opposite in Ba$_2$IrO$_4$ and Sr$_2$IrO$_4$, despite the fact that both materials display a similar elongation of IrO$_6$ octahedra.  Since this octahedral distortion is expected to generate a positive tetragonal crystal field splitting in both systems, the apparent reversal of $j_{eff}$ = 3/2 levels demonstrates that longer-range crystal field effects must play a significant role in these materials as pointed out in Ref.~\onlinecite{Bogdanov2015}.  This reversal of energy levels is consistent with recent x-ray absorption measurements on Ba$_2$IrO$_4$\cite{MorettiSala2014c} and Sr$_2$IrO$_4$\cite{Haskel2012}.

A comparison of excitation spectra for Ba$_2$IrO$_4$ (glancing geometry) and Sr$_2$IrO$_4$ (normal geometry, as in Ref.~\onlinecite{JKim2014}) also reveals a clear suppression of the spin-orbit exciton energy scale.  The sharp exciton feature at {\bf Q} = ($\pi$/2, $\pi$/2) shifts from $\sim$470 meV in Sr$_2$IrO$_4$ to $\sim$350 meV in Ba$_2$IrO$_4$.  This trend is opposite to predictions from quantum chemistry calculations\cite{Katukuri2012}, and approximately four times larger than predicted by theory. This peak shift seems to be due to the larger bandwidth of the spin-orbit exciton in Ba$_2$IrO$_4$, as the zone center energy of $\sim 600$~meV is similar to that in Sr$_2$IrO$_4$ (See Fig. 3(d)). We note that the observed bandwidth of $\sim$250~meV in Ba$_2$IrO$_4$ is almost a factor of two larger than that of Sr$_2$IrO$_4$. However, the apparent larger bandwidth could be due to the presence of two spin-orbit exciton modes and a further investigation of the incident angle dependence is necessary to disentangle these two modes, as was done for Sr$_2$IrO$_4$.~\cite{JKim2014}

The strain dependence of the spin-orbit exciton mode (Fig.~\ref{fig_rixs}(c,d)) suggests that epitaxial strain can also be used to tune the orbital properties of Ba$_2$IrO$_4$.  As the compressive strain increases, we expect the {\it c}/{\it a}-ratio to grow larger and the elongation of the IrO$_6$ octahedra to become more pronounced.  This octahedral distortion should make the non-cubic crystal field splitting, $\Delta_{CEF}$, larger in Ba$_2$IrO$_4$ (See Fig.~\ref{fig_comp}(b)).  In principle, this suggests that it should be possible to tune Ba$_2$IrO$_4$ (or Sr$_2$IrO$_4$) to the $\Delta_{CEF}$ = 0 limit where the magnetic anisotropy vanishes, by applying an appropriate epitaxial strain.  In the case of Ba$_2$IrO$_4$ the required strain should be tensile (to decrease $\Delta_{CEF}$), while for Sr$_2$IrO$_4$ it should be compressive (to increase $\Delta_{CEF}$).  This conclusion is supported by recent LDA + DMFT calculations on Sr$_2$IrO$_4$\cite{Zhang2013}, which predict that a compressive strain of $\sim$1\% would be required to tune the local structure to the $\Delta_{CEF}$ = 0 point.  It is interesting to note that the $\Delta_{CEF}$ = 0 limit in Ba$_2$IrO$_4$ and Sr$_2$IrO$_4$, which restores full rotational symmetry, does not correspond to perfectly undistorted IrO$_6$ octahedra, but rather to the point where short-range and long-range crystal field effects cancel out.

\section{Summary and Conclusions}

In summary, our RIXS studies of ultra-thin epitaxial films of Ba$_2$IrO$_4$ reveal that the bandwidths of the low-energy magnetic and spin-orbit excitations in this material are significantly larger than those found in Sr$_2$IrO$_4$. This observation can be understood from the increased hopping due to the 180$^\circ$ Ir-O-Ir bond in Ba$_2$IrO$_4$, which suggests that itinerancy may be important for understanding the electronic structure of Ba$_2$IrO$_4$. The results of this study demonstrate that hard x-ray RIXS can provide a detailed probe of spin and orbital excitations in ultrathin film samples.  This technique offers unique opportunities for the study of samples which are difficult, or impossible, to synthesize in bulk single crystal form, allowing novel materials to be investigated under a variety of highly tuneable strain conditions (tensile or compressive, uniform or anisotropic).  In closing, we note that the relatively high inelastic count rates in this study ($\sim$5 cts/s) resulted in very reasonable measurement times ($\sim$1 to 2 min/pt for the data in Figures 3 and 4).  This suggests that measurements on monolayer and bilayer thin films are entirely feasible even with hard x-rays, opening up exciting new possibilities for the study of multilayers, buried interfaces, and complex heterostructures.

\section*{Acknowledgement}

We thank J. van den Brink for invaluable discussions. Work at the University of Toronto was supported by the Natural Sciences and Engineering Research Council (NSERC) of Canada, the Banting Postdoctoral Fellowship program, and the Canada Research Chairs program. Work at Cornell University was supported by the Air Force Office of Scientific Research (Grant No. FA9550-21-1-0168), and the National Science Foundation (No. DMR-2104427). M.U. acknowledges the support by the JSPS Postdoctoral Fellowships for Research Abroad. Use of the Advanced Photon Source at Argonne National Laboratory is supported by the U.S. Department of Energy, Office of Science, Office of Basic Energy Sciences, under Contract No. DE-AC02-06CH11357.

\bibliography{Ba214}

\begin{thebibliography}{32}
\expandafter\ifx\csname natexlab\endcsname\relax\def\natexlab#1{#1}\fi
\expandafter\ifx\csname bibnamefont\endcsname\relax
  \def\bibnamefont#1{#1}\fi
\expandafter\ifx\csname bibfnamefont\endcsname\relax
  \def\bibfnamefont#1{#1}\fi
\expandafter\ifx\csname citenamefont\endcsname\relax
  \def\citenamefont#1{#1}\fi
\expandafter\ifx\csname url\endcsname\relax
  \def\url#1{\texttt{#1}}\fi
\expandafter\ifx\csname urlprefix\endcsname\relax\def\urlprefix{URL }\fi
\providecommand{\bibinfo}[2]{#2}
\providecommand{\eprint}[2][]{\url{#2}}

\bibitem[{\citenamefont{Kim et~al.}(2008)\citenamefont{Kim, Jin, Moon, Kim,
  Park, Leem, Yu, Noh, Kim, Oh et~al.}}]{BJKim2008}
\bibinfo{author}{\bibfnamefont{B.~J.} \bibnamefont{Kim}},
  \bibinfo{author}{\bibfnamefont{H.}~\bibnamefont{Jin}},
  \bibinfo{author}{\bibfnamefont{S.~J.} \bibnamefont{Moon}},
  \bibinfo{author}{\bibfnamefont{J.-Y.} \bibnamefont{Kim}},
  \bibinfo{author}{\bibfnamefont{B.-G.} \bibnamefont{Park}},
  \bibinfo{author}{\bibfnamefont{C.~S.} \bibnamefont{Leem}},
  \bibinfo{author}{\bibfnamefont{J.}~\bibnamefont{Yu}},
  \bibinfo{author}{\bibfnamefont{T.~W.} \bibnamefont{Noh}},
  \bibinfo{author}{\bibfnamefont{C.}~\bibnamefont{Kim}},
  \bibinfo{author}{\bibfnamefont{S.-J.} \bibnamefont{Oh}},
  \bibnamefont{et~al.}, \bibinfo{journal}{Phys. Rev. Lett.}
  \textbf{\bibinfo{volume}{101}}, \bibinfo{pages}{076402}
  (\bibinfo{year}{2008}).

\bibitem[{\citenamefont{Kim et~al.}(2009)\citenamefont{Kim, Ohsumi, Komesu,
  Sakai, Morita, Takagi, and Arima}}]{BJKim2009}
\bibinfo{author}{\bibfnamefont{B.~J.} \bibnamefont{Kim}},
  \bibinfo{author}{\bibfnamefont{H.}~\bibnamefont{Ohsumi}},
  \bibinfo{author}{\bibfnamefont{T.}~\bibnamefont{Komesu}},
  \bibinfo{author}{\bibfnamefont{S.}~\bibnamefont{Sakai}},
  \bibinfo{author}{\bibfnamefont{T.}~\bibnamefont{Morita}},
  \bibinfo{author}{\bibfnamefont{H.}~\bibnamefont{Takagi}}, \bibnamefont{and}
  \bibinfo{author}{\bibfnamefont{T.}~\bibnamefont{Arima}},
  \bibinfo{journal}{Science} \textbf{\bibinfo{volume}{323}},
  \bibinfo{pages}{1329} (\bibinfo{year}{2009}),
  \urlprefix\url{http://www.sciencemag.org/content/323/5919/1329.abstract}.

\bibitem[{\citenamefont{Jackeli and Khaliullin}(2009)}]{Jackeli2009}
\bibinfo{author}{\bibfnamefont{G.}~\bibnamefont{Jackeli}} \bibnamefont{and}
  \bibinfo{author}{\bibfnamefont{G.}~\bibnamefont{Khaliullin}},
  \bibinfo{journal}{Phys. Rev. Lett.} \textbf{\bibinfo{volume}{102}},
  \bibinfo{pages}{017205} (\bibinfo{year}{2009}).

\bibitem[{\citenamefont{Cao and Schlottmann}(2018)}]{Cao2018}
\bibinfo{author}{\bibfnamefont{G.}~\bibnamefont{Cao}} \bibnamefont{and}
  \bibinfo{author}{\bibfnamefont{P.}~\bibnamefont{Schlottmann}},
  \bibinfo{journal}{Rep. Prog. Phys.} \textbf{\bibinfo{volume}{81}},
  \bibinfo{pages}{042502} (\bibinfo{year}{2018}), ISSN
  \bibinfo{issn}{0034-4885}, \bibinfo{note}{publisher: IOP Publishing},
  \urlprefix\url{https://doi.org/10.1088/1361-6633/aaa979}.

\bibitem[{\citenamefont{Bertinshaw et~al.}(2019)\citenamefont{Bertinshaw, Kim,
  Khaliullin, and Kim}}]{Bertinshaw2019}
\bibinfo{author}{\bibfnamefont{J.}~\bibnamefont{Bertinshaw}},
  \bibinfo{author}{\bibfnamefont{Y.}~\bibnamefont{Kim}},
  \bibinfo{author}{\bibfnamefont{G.}~\bibnamefont{Khaliullin}},
  \bibnamefont{and} \bibinfo{author}{\bibfnamefont{B.}~\bibnamefont{Kim}},
  \bibinfo{journal}{Annual Review of Condensed Matter Physics}
  \textbf{\bibinfo{volume}{10}}, \bibinfo{pages}{315} (\bibinfo{year}{2019}),
  \eprint{https://doi.org/10.1146/annurev-conmatphys-031218-013113},
  \urlprefix\url{https://doi.org/10.1146/annurev-conmatphys-031218-013113}.

\bibitem[{\citenamefont{Wang and Senthil}(2011)}]{Wang2011}
\bibinfo{author}{\bibfnamefont{F.}~\bibnamefont{Wang}} \bibnamefont{and}
  \bibinfo{author}{\bibfnamefont{T.}~\bibnamefont{Senthil}},
  \bibinfo{journal}{Phys. Rev. Lett.} \textbf{\bibinfo{volume}{106}},
  \bibinfo{pages}{136402} (\bibinfo{year}{2011}),
  \urlprefix\url{https://link.aps.org/doi/10.1103/PhysRevLett.106.136402}.

\bibitem[{\citenamefont{Watanabe et~al.}(2013)\citenamefont{Watanabe,
  Shirakawa, and Yunoki}}]{Watanabe2013}
\bibinfo{author}{\bibfnamefont{H.}~\bibnamefont{Watanabe}},
  \bibinfo{author}{\bibfnamefont{T.}~\bibnamefont{Shirakawa}},
  \bibnamefont{and} \bibinfo{author}{\bibfnamefont{S.}~\bibnamefont{Yunoki}},
  \bibinfo{journal}{Phys. Rev. Lett.} \textbf{\bibinfo{volume}{110}},
  \bibinfo{pages}{027002} (\bibinfo{year}{2013}).

\bibitem[{\citenamefont{Kim et~al.}(2012)\citenamefont{Kim, Casa, Upton, Gog,
  Kim, Mitchell, van Veenendaal, Daghofer, van~den Brink, Khaliullin
  et~al.}}]{JKim2012a}
\bibinfo{author}{\bibfnamefont{J.}~\bibnamefont{Kim}},
  \bibinfo{author}{\bibfnamefont{D.}~\bibnamefont{Casa}},
  \bibinfo{author}{\bibfnamefont{M.~H.} \bibnamefont{Upton}},
  \bibinfo{author}{\bibfnamefont{T.}~\bibnamefont{Gog}},
  \bibinfo{author}{\bibfnamefont{Y.-J.} \bibnamefont{Kim}},
  \bibinfo{author}{\bibfnamefont{J.~F.} \bibnamefont{Mitchell}},
  \bibinfo{author}{\bibfnamefont{M.}~\bibnamefont{van Veenendaal}},
  \bibinfo{author}{\bibfnamefont{M.}~\bibnamefont{Daghofer}},
  \bibinfo{author}{\bibfnamefont{J.}~\bibnamefont{van~den Brink}},
  \bibinfo{author}{\bibfnamefont{G.}~\bibnamefont{Khaliullin}},
  \bibnamefont{et~al.}, \bibinfo{journal}{Phys. Rev. Lett.}
  \textbf{\bibinfo{volume}{108}}, \bibinfo{pages}{177003}
  (\bibinfo{year}{2012}).

\bibitem[{\citenamefont{Yang et~al.}(2014)\citenamefont{Yang, Wang, Liu, Chen,
  Dai, and Wang}}]{Yang2014}
\bibinfo{author}{\bibfnamefont{Y.}~\bibnamefont{Yang}},
  \bibinfo{author}{\bibfnamefont{W.-S.} \bibnamefont{Wang}},
  \bibinfo{author}{\bibfnamefont{J.-G.} \bibnamefont{Liu}},
  \bibinfo{author}{\bibfnamefont{H.}~\bibnamefont{Chen}},
  \bibinfo{author}{\bibfnamefont{J.-H.} \bibnamefont{Dai}}, \bibnamefont{and}
  \bibinfo{author}{\bibfnamefont{Q.-H.} \bibnamefont{Wang}},
  \bibinfo{journal}{Phys. Rev. B} \textbf{\bibinfo{volume}{89}},
  \bibinfo{pages}{094518} (\bibinfo{year}{2014}),
  \urlprefix\url{https://link.aps.org/doi/10.1103/PhysRevB.89.094518}.

\bibitem[{\citenamefont{Okabe et~al.}(2011)\citenamefont{Okabe, Isobe,
  Takayama-Muromachi, Koda, Takeshita, Hiraishi, Miyazaki, Kadono, Miyake, and
  Akimitsu}}]{Okabe2011}
\bibinfo{author}{\bibfnamefont{H.}~\bibnamefont{Okabe}},
  \bibinfo{author}{\bibfnamefont{M.}~\bibnamefont{Isobe}},
  \bibinfo{author}{\bibfnamefont{E.}~\bibnamefont{Takayama-Muromachi}},
  \bibinfo{author}{\bibfnamefont{A.}~\bibnamefont{Koda}},
  \bibinfo{author}{\bibfnamefont{S.}~\bibnamefont{Takeshita}},
  \bibinfo{author}{\bibfnamefont{M.}~\bibnamefont{Hiraishi}},
  \bibinfo{author}{\bibfnamefont{M.}~\bibnamefont{Miyazaki}},
  \bibinfo{author}{\bibfnamefont{R.}~\bibnamefont{Kadono}},
  \bibinfo{author}{\bibfnamefont{Y.}~\bibnamefont{Miyake}}, \bibnamefont{and}
  \bibinfo{author}{\bibfnamefont{J.}~\bibnamefont{Akimitsu}},
  \bibinfo{journal}{Phys. Rev. B} \textbf{\bibinfo{volume}{83}},
  \bibinfo{pages}{155118} (\bibinfo{year}{2011}),
  \urlprefix\url{https://link.aps.org/doi/10.1103/PhysRevB.83.155118}.

\bibitem[{\citenamefont{Boseggia et~al.}(2013)\citenamefont{Boseggia,
  Springell, Walker, R\o{}nnow, R\"uegg, Okabe, Isobe, Perry, Collins, and
  McMorrow}}]{Boseggia2013}
\bibinfo{author}{\bibfnamefont{S.}~\bibnamefont{Boseggia}},
  \bibinfo{author}{\bibfnamefont{R.}~\bibnamefont{Springell}},
  \bibinfo{author}{\bibfnamefont{H.~C.} \bibnamefont{Walker}},
  \bibinfo{author}{\bibfnamefont{H.~M.} \bibnamefont{R\o{}nnow}},
  \bibinfo{author}{\bibfnamefont{C.}~\bibnamefont{R\"uegg}},
  \bibinfo{author}{\bibfnamefont{H.}~\bibnamefont{Okabe}},
  \bibinfo{author}{\bibfnamefont{M.}~\bibnamefont{Isobe}},
  \bibinfo{author}{\bibfnamefont{R.~S.} \bibnamefont{Perry}},
  \bibinfo{author}{\bibfnamefont{S.~P.} \bibnamefont{Collins}},
  \bibnamefont{and} \bibinfo{author}{\bibfnamefont{D.~F.}
  \bibnamefont{McMorrow}}, \bibinfo{journal}{Phys. Rev. Lett.}
  \textbf{\bibinfo{volume}{110}}, \bibinfo{pages}{117207}
  (\bibinfo{year}{2013}),
  \urlprefix\url{https://link.aps.org/doi/10.1103/PhysRevLett.110.117207}.

\bibitem[{\citenamefont{Moser et~al.}(2014)\citenamefont{Moser, Moreschini,
  Ebrahimi, Piazza, Isobe, Okabe, Akimitsu, Mazurenko, Kim, Bostwick
  et~al.}}]{Moser2014}
\bibinfo{author}{\bibfnamefont{S.}~\bibnamefont{Moser}},
  \bibinfo{author}{\bibfnamefont{L.}~\bibnamefont{Moreschini}},
  \bibinfo{author}{\bibfnamefont{A.}~\bibnamefont{Ebrahimi}},
  \bibinfo{author}{\bibfnamefont{B.~D.} \bibnamefont{Piazza}},
  \bibinfo{author}{\bibfnamefont{M.}~\bibnamefont{Isobe}},
  \bibinfo{author}{\bibfnamefont{H.}~\bibnamefont{Okabe}},
  \bibinfo{author}{\bibfnamefont{J.}~\bibnamefont{Akimitsu}},
  \bibinfo{author}{\bibfnamefont{V.~V.} \bibnamefont{Mazurenko}},
  \bibinfo{author}{\bibfnamefont{K.~S.} \bibnamefont{Kim}},
  \bibinfo{author}{\bibfnamefont{A.}~\bibnamefont{Bostwick}},
  \bibnamefont{et~al.}, \bibinfo{journal}{New Journal of Physics}
  \textbf{\bibinfo{volume}{16}}, \bibinfo{pages}{013008}
  (\bibinfo{year}{2014}),
  \urlprefix\url{http://stacks.iop.org/1367-2630/16/i=1/a=013008}.

\bibitem[{\citenamefont{Moretti~Sala
  et~al.}(2014{\natexlab{a}})\citenamefont{Moretti~Sala, Rossi, Al-Zein,
  Boseggia, Hunter, Perry, Prabhakaran, Boothroyd, Brookes, McMorrow
  et~al.}}]{MorettiSala2014b}
\bibinfo{author}{\bibfnamefont{M.}~\bibnamefont{Moretti~Sala}},
  \bibinfo{author}{\bibfnamefont{M.}~\bibnamefont{Rossi}},
  \bibinfo{author}{\bibfnamefont{A.}~\bibnamefont{Al-Zein}},
  \bibinfo{author}{\bibfnamefont{S.}~\bibnamefont{Boseggia}},
  \bibinfo{author}{\bibfnamefont{E.~C.} \bibnamefont{Hunter}},
  \bibinfo{author}{\bibfnamefont{R.~S.} \bibnamefont{Perry}},
  \bibinfo{author}{\bibfnamefont{D.}~\bibnamefont{Prabhakaran}},
  \bibinfo{author}{\bibfnamefont{A.~T.} \bibnamefont{Boothroyd}},
  \bibinfo{author}{\bibfnamefont{N.~B.} \bibnamefont{Brookes}},
  \bibinfo{author}{\bibfnamefont{D.~F.} \bibnamefont{McMorrow}},
  \bibnamefont{et~al.}, \bibinfo{journal}{Phys. Rev. B}
  \textbf{\bibinfo{volume}{90}}, \bibinfo{pages}{085126}
  (\bibinfo{year}{2014}{\natexlab{a}}),
  \urlprefix\url{http://link.aps.org/doi/10.1103/PhysRevB.90.085126}.

\bibitem[{\citenamefont{Uchida et~al.}(2014)\citenamefont{Uchida, Nie, King,
  Kim, Fennie, Schlom, and Shen}}]{Uchida2014}
\bibinfo{author}{\bibfnamefont{M.}~\bibnamefont{Uchida}},
  \bibinfo{author}{\bibfnamefont{Y.~F.} \bibnamefont{Nie}},
  \bibinfo{author}{\bibfnamefont{P.~D.~C.} \bibnamefont{King}},
  \bibinfo{author}{\bibfnamefont{C.~H.} \bibnamefont{Kim}},
  \bibinfo{author}{\bibfnamefont{C.~J.} \bibnamefont{Fennie}},
  \bibinfo{author}{\bibfnamefont{D.~G.} \bibnamefont{Schlom}},
  \bibnamefont{and} \bibinfo{author}{\bibfnamefont{K.~M.} \bibnamefont{Shen}},
  \bibinfo{journal}{Phys. Rev. B} \textbf{\bibinfo{volume}{90}},
  \bibinfo{pages}{075142} (\bibinfo{year}{2014}),
  \urlprefix\url{https://link.aps.org/doi/10.1103/PhysRevB.90.075142}.

\bibitem[{\citenamefont{Solovyev et~al.}(2015)\citenamefont{Solovyev,
  Mazurenko, and Katanin}}]{Solovyev2015}
\bibinfo{author}{\bibfnamefont{I.~V.} \bibnamefont{Solovyev}},
  \bibinfo{author}{\bibfnamefont{V.~V.} \bibnamefont{Mazurenko}},
  \bibnamefont{and} \bibinfo{author}{\bibfnamefont{A.~A.}
  \bibnamefont{Katanin}}, \bibinfo{journal}{Phys. Rev. B}
  \textbf{\bibinfo{volume}{92}}, \bibinfo{pages}{235109}
  (\bibinfo{year}{2015}),
  \urlprefix\url{https://link.aps.org/doi/10.1103/PhysRevB.92.235109}.

\bibitem[{\citenamefont{Ro\ifmmode~\acute{s}\else \'{s}\fi{}ciszewski and
  Ole\ifmmode~\acute{s}\else \'{s}\fi{}}(2016)}]{Rosciszewski2016}
\bibinfo{author}{\bibfnamefont{K.}~\bibnamefont{Ro\ifmmode~\acute{s}\else
  \'{s}\fi{}ciszewski}} \bibnamefont{and} \bibinfo{author}{\bibfnamefont{A.~M.}
  \bibnamefont{Ole\ifmmode~\acute{s}\else \'{s}\fi{}}}, \bibinfo{journal}{Phys.
  Rev. B} \textbf{\bibinfo{volume}{93}}, \bibinfo{pages}{085106}
  (\bibinfo{year}{2016}),
  \urlprefix\url{https://link.aps.org/doi/10.1103/PhysRevB.93.085106}.

\bibitem[{\citenamefont{Katukuri et~al.}(2012)\citenamefont{Katukuri, Stoll,
  van~den Brink, and Hozoi}}]{Katukuri2012}
\bibinfo{author}{\bibfnamefont{V.~M.} \bibnamefont{Katukuri}},
  \bibinfo{author}{\bibfnamefont{H.}~\bibnamefont{Stoll}},
  \bibinfo{author}{\bibfnamefont{J.}~\bibnamefont{van~den Brink}},
  \bibnamefont{and} \bibinfo{author}{\bibfnamefont{L.}~\bibnamefont{Hozoi}},
  \bibinfo{journal}{Phys. Rev. B} \textbf{\bibinfo{volume}{85}},
  \bibinfo{pages}{220402} (\bibinfo{year}{2012}),
  \urlprefix\url{http://link.aps.org/doi/10.1103/PhysRevB.85.220402}.

\bibitem[{\citenamefont{Tsuda et~al.}(2016)\citenamefont{Tsuda, Okabe, Isobe,
  and Uji}}]{Tsuda2016}
\bibinfo{author}{\bibfnamefont{S.}~\bibnamefont{Tsuda}},
  \bibinfo{author}{\bibfnamefont{H.}~\bibnamefont{Okabe}},
  \bibinfo{author}{\bibfnamefont{M.}~\bibnamefont{Isobe}}, \bibnamefont{and}
  \bibinfo{author}{\bibfnamefont{S.}~\bibnamefont{Uji}}, \bibinfo{journal}{J.
  Phys. Soc. Jpn.} \textbf{\bibinfo{volume}{85}}, \bibinfo{pages}{023703}
  (\bibinfo{year}{2016}), ISSN \bibinfo{issn}{0031-9015},
  \bibinfo{note}{publisher: The Physical Society of Japan},
  \urlprefix\url{https://journals.jps.jp/doi/10.7566/JPSJ.85.023703}.

\bibitem[{\citenamefont{Bogdanov et~al.}(2015)\citenamefont{Bogdanov, Katukuri,
  Romhanyi, Yushankhai, Kataev, B\"uchner, van~den Brink, and
  Hozoi}}]{Bogdanov2015}
\bibinfo{author}{\bibfnamefont{N.~A.} \bibnamefont{Bogdanov}},
  \bibinfo{author}{\bibfnamefont{V.~M.} \bibnamefont{Katukuri}},
  \bibinfo{author}{\bibfnamefont{J.}~\bibnamefont{Romhanyi}},
  \bibinfo{author}{\bibfnamefont{V.}~\bibnamefont{Yushankhai}},
  \bibinfo{author}{\bibfnamefont{V.}~\bibnamefont{Kataev}},
  \bibinfo{author}{\bibfnamefont{B.}~\bibnamefont{B\"uchner}},
  \bibinfo{author}{\bibfnamefont{J.}~\bibnamefont{van~den Brink}},
  \bibnamefont{and} \bibinfo{author}{\bibfnamefont{L.}~\bibnamefont{Hozoi}},
  \bibinfo{journal}{Nat. Comm.} \textbf{\bibinfo{volume}{6}},
  \bibinfo{pages}{7306} (\bibinfo{year}{2015}).

\bibitem[{\citenamefont{Katukuri et~al.}(2014)\citenamefont{Katukuri,
  Yushankhai, Siurakshina, van~den Brink, Hozoi, and
  Rousochatzakis}}]{Katukuri2014c}
\bibinfo{author}{\bibfnamefont{V.~M.} \bibnamefont{Katukuri}},
  \bibinfo{author}{\bibfnamefont{V.}~\bibnamefont{Yushankhai}},
  \bibinfo{author}{\bibfnamefont{L.}~\bibnamefont{Siurakshina}},
  \bibinfo{author}{\bibfnamefont{J.}~\bibnamefont{van~den Brink}},
  \bibinfo{author}{\bibfnamefont{L.}~\bibnamefont{Hozoi}}, \bibnamefont{and}
  \bibinfo{author}{\bibfnamefont{I.}~\bibnamefont{Rousochatzakis}},
  \bibinfo{journal}{Phys. Rev. X} \textbf{\bibinfo{volume}{4}},
  \bibinfo{pages}{021051} (\bibinfo{year}{2014}),
  \urlprefix\url{https://link.aps.org/doi/10.1103/PhysRevX.4.021051}.

\bibitem[{\citenamefont{Hou et~al.}(2016)\citenamefont{Hou, Xiang, and
  Gong}}]{Hou2016}
\bibinfo{author}{\bibfnamefont{Y.~S.} \bibnamefont{Hou}},
  \bibinfo{author}{\bibfnamefont{H.~J.} \bibnamefont{Xiang}}, \bibnamefont{and}
  \bibinfo{author}{\bibfnamefont{X.~G.} \bibnamefont{Gong}},
  \bibinfo{journal}{New Journal of Physics} \textbf{\bibinfo{volume}{18}},
  \bibinfo{pages}{043007} (\bibinfo{year}{2016}),
  \urlprefix\url{https://doi.org/10.1088/1367-2630/18/4/043007}.

\bibitem[{\citenamefont{Nichols et~al.}(2014)\citenamefont{Nichols, Korneta,
  Terzic, Cao, Brill, and Seo}}]{Nichols2014}
\bibinfo{author}{\bibfnamefont{J.}~\bibnamefont{Nichols}},
  \bibinfo{author}{\bibfnamefont{O.~B.} \bibnamefont{Korneta}},
  \bibinfo{author}{\bibfnamefont{J.}~\bibnamefont{Terzic}},
  \bibinfo{author}{\bibfnamefont{G.}~\bibnamefont{Cao}},
  \bibinfo{author}{\bibfnamefont{J.~W.} \bibnamefont{Brill}}, \bibnamefont{and}
  \bibinfo{author}{\bibfnamefont{S.~S.~A.} \bibnamefont{Seo}},
  \bibinfo{journal}{Applied Physics Letters} \textbf{\bibinfo{volume}{104}},
  \bibinfo{pages}{121913} (\bibinfo{year}{2014}),
  \eprint{https://doi.org/10.1063/1.4870049},
  \urlprefix\url{https://doi.org/10.1063/1.4870049}.

\bibitem[{\citenamefont{Zhao et~al.}(2021)\citenamefont{Zhao, Zhang, Cai, Guo,
  Ji, Zhang, Gu, Zhou, Zhu, and Nie}}]{Zhao2021}
\bibinfo{author}{\bibfnamefont{Y.-Q.} \bibnamefont{Zhao}},
  \bibinfo{author}{\bibfnamefont{H.}~\bibnamefont{Zhang}},
  \bibinfo{author}{\bibfnamefont{X.-B.} \bibnamefont{Cai}},
  \bibinfo{author}{\bibfnamefont{W.}~\bibnamefont{Guo}},
  \bibinfo{author}{\bibfnamefont{D.-X.} \bibnamefont{Ji}},
  \bibinfo{author}{\bibfnamefont{T.-T.} \bibnamefont{Zhang}},
  \bibinfo{author}{\bibfnamefont{Z.-B.} \bibnamefont{Gu}},
  \bibinfo{author}{\bibfnamefont{J.}~\bibnamefont{Zhou}},
  \bibinfo{author}{\bibfnamefont{Y.}~\bibnamefont{Zhu}}, \bibnamefont{and}
  \bibinfo{author}{\bibfnamefont{Y.-F.} \bibnamefont{Nie}},
  \bibinfo{journal}{Chinese Phys. B} \textbf{\bibinfo{volume}{30}},
  \bibinfo{pages}{087401} (\bibinfo{year}{2021}), ISSN
  \bibinfo{issn}{1674-1056}, \bibinfo{note}{publisher: IOP Publishing},
  \urlprefix\url{https://doi.org/10.1088/1674-1056/abea97}.

\bibitem[{\citenamefont{Braicovich et~al.}(2009)\citenamefont{Braicovich,
  Ament, Bisogni, Forte, Aruta, Balestrino, Brookes, De~Luca, Medaglia,
  Granozio et~al.}}]{Braicovich2009}
\bibinfo{author}{\bibfnamefont{L.}~\bibnamefont{Braicovich}},
  \bibinfo{author}{\bibfnamefont{L.~J.~P.} \bibnamefont{Ament}},
  \bibinfo{author}{\bibfnamefont{V.}~\bibnamefont{Bisogni}},
  \bibinfo{author}{\bibfnamefont{F.}~\bibnamefont{Forte}},
  \bibinfo{author}{\bibfnamefont{C.}~\bibnamefont{Aruta}},
  \bibinfo{author}{\bibfnamefont{G.}~\bibnamefont{Balestrino}},
  \bibinfo{author}{\bibfnamefont{N.~B.} \bibnamefont{Brookes}},
  \bibinfo{author}{\bibfnamefont{G.~M.} \bibnamefont{De~Luca}},
  \bibinfo{author}{\bibfnamefont{P.~G.} \bibnamefont{Medaglia}},
  \bibinfo{author}{\bibfnamefont{F.~M.} \bibnamefont{Granozio}},
  \bibnamefont{et~al.}, \bibinfo{journal}{Phys. Rev. Lett.}
  \textbf{\bibinfo{volume}{102}}, \bibinfo{pages}{167401}
  (\bibinfo{year}{2009}),
  \urlprefix\url{http://link.aps.org/doi/10.1103/PhysRevLett.102.167401}.

\bibitem[{\citenamefont{Schlappa et~al.}(2012)\citenamefont{Schlappa, Wohlfeld,
  Zhou, Mourigal, Haverkort, Strocov, Hozoi, Monney, Nishimoto, Singh
  et~al.}}]{Schlappa2012}
\bibinfo{author}{\bibfnamefont{J.}~\bibnamefont{Schlappa}},
  \bibinfo{author}{\bibfnamefont{K.}~\bibnamefont{Wohlfeld}},
  \bibinfo{author}{\bibfnamefont{K.}~\bibnamefont{Zhou}},
  \bibinfo{author}{\bibfnamefont{M.}~\bibnamefont{Mourigal}},
  \bibinfo{author}{\bibfnamefont{M.}~\bibnamefont{Haverkort}},
  \bibinfo{author}{\bibfnamefont{V.}~\bibnamefont{Strocov}},
  \bibinfo{author}{\bibfnamefont{L.}~\bibnamefont{Hozoi}},
  \bibinfo{author}{\bibfnamefont{C.}~\bibnamefont{Monney}},
  \bibinfo{author}{\bibfnamefont{S.}~\bibnamefont{Nishimoto}},
  \bibinfo{author}{\bibfnamefont{S.}~\bibnamefont{Singh}},
  \bibnamefont{et~al.}, \bibinfo{journal}{Nature}
  \textbf{\bibinfo{volume}{485}}, \bibinfo{pages}{82} (\bibinfo{year}{2012}).

\bibitem[{\citenamefont{Cao et~al.}(1998)\citenamefont{Cao, Bolivar, McCall,
  Crow, and Guertin}}]{Cao1998}
\bibinfo{author}{\bibfnamefont{G.}~\bibnamefont{Cao}},
  \bibinfo{author}{\bibfnamefont{J.}~\bibnamefont{Bolivar}},
  \bibinfo{author}{\bibfnamefont{S.}~\bibnamefont{McCall}},
  \bibinfo{author}{\bibfnamefont{J.~E.} \bibnamefont{Crow}}, \bibnamefont{and}
  \bibinfo{author}{\bibfnamefont{R.~P.} \bibnamefont{Guertin}},
  \bibinfo{journal}{Phys. Rev. B} \textbf{\bibinfo{volume}{57}},
  \bibinfo{pages}{R11039} (\bibinfo{year}{1998}),
  \urlprefix\url{http://link.aps.org/doi/10.1103/PhysRevB.57.R11039}.

\bibitem[{\citenamefont{Lupascu et~al.}(2014)\citenamefont{Lupascu, Clancy,
  Gretarsson, Nie, Nichols, Terzic, Cao, Seo, Islam, Upton
  et~al.}}]{Lupascu2014}
\bibinfo{author}{\bibfnamefont{A.}~\bibnamefont{Lupascu}},
  \bibinfo{author}{\bibfnamefont{J.~P.} \bibnamefont{Clancy}},
  \bibinfo{author}{\bibfnamefont{H.}~\bibnamefont{Gretarsson}},
  \bibinfo{author}{\bibfnamefont{Z.}~\bibnamefont{Nie}},
  \bibinfo{author}{\bibfnamefont{J.}~\bibnamefont{Nichols}},
  \bibinfo{author}{\bibfnamefont{J.}~\bibnamefont{Terzic}},
  \bibinfo{author}{\bibfnamefont{G.}~\bibnamefont{Cao}},
  \bibinfo{author}{\bibfnamefont{S.~S.~A.} \bibnamefont{Seo}},
  \bibinfo{author}{\bibfnamefont{Z.}~\bibnamefont{Islam}},
  \bibinfo{author}{\bibfnamefont{M.~H.} \bibnamefont{Upton}},
  \bibnamefont{et~al.}, \bibinfo{journal}{Physical Review Letters}
  \textbf{\bibinfo{volume}{112}} (\bibinfo{year}{2014}), ISSN
  \bibinfo{issn}{0031-9007}, \urlprefix\url{<Go to ISI>://WOS:000339490700015}.

\bibitem[{\citenamefont{Coldea et~al.}(2001)\citenamefont{Coldea, Hayden,
  Aeppli, Perring, Frost, Mason, Cheong, and Fisk}}]{Coldea2001}
\bibinfo{author}{\bibfnamefont{R.}~\bibnamefont{Coldea}},
  \bibinfo{author}{\bibfnamefont{S.~M.} \bibnamefont{Hayden}},
  \bibinfo{author}{\bibfnamefont{G.}~\bibnamefont{Aeppli}},
  \bibinfo{author}{\bibfnamefont{T.~G.} \bibnamefont{Perring}},
  \bibinfo{author}{\bibfnamefont{C.~D.} \bibnamefont{Frost}},
  \bibinfo{author}{\bibfnamefont{T.~E.} \bibnamefont{Mason}},
  \bibinfo{author}{\bibfnamefont{S.-W.} \bibnamefont{Cheong}},
  \bibnamefont{and} \bibinfo{author}{\bibfnamefont{Z.}~\bibnamefont{Fisk}},
  \bibinfo{journal}{Phys. Rev. Lett.} \textbf{\bibinfo{volume}{86}},
  \bibinfo{pages}{5377} (\bibinfo{year}{2001}),
  \urlprefix\url{https://link.aps.org/doi/10.1103/PhysRevLett.86.5377}.

\bibitem[{\citenamefont{Kim et~al.}(2014)\citenamefont{Kim, Daghofer, Said,
  Gog, van~den Brink, Khaliullin, and Kim}}]{JKim2014}
\bibinfo{author}{\bibfnamefont{J.}~\bibnamefont{Kim}},
  \bibinfo{author}{\bibfnamefont{M.}~\bibnamefont{Daghofer}},
  \bibinfo{author}{\bibfnamefont{A.~H.} \bibnamefont{Said}},
  \bibinfo{author}{\bibfnamefont{T.}~\bibnamefont{Gog}},
  \bibinfo{author}{\bibfnamefont{J.}~\bibnamefont{van~den Brink}},
  \bibinfo{author}{\bibfnamefont{G.}~\bibnamefont{Khaliullin}},
  \bibnamefont{and} \bibinfo{author}{\bibfnamefont{B.~J.} \bibnamefont{Kim}},
  \bibinfo{journal}{Nat. Comm.} \textbf{\bibinfo{volume}{5}},
  \bibinfo{pages}{4453} (\bibinfo{year}{2014}).

\bibitem[{\citenamefont{Moretti~Sala
  et~al.}(2014{\natexlab{b}})\citenamefont{Moretti~Sala, Rossi, Boseggia,
  Akimitsu, Brookes, Isobe, Minola, Okabe, R\o{}nnow, Simonelli
  et~al.}}]{MorettiSala2014c}
\bibinfo{author}{\bibfnamefont{M.}~\bibnamefont{Moretti~Sala}},
  \bibinfo{author}{\bibfnamefont{M.}~\bibnamefont{Rossi}},
  \bibinfo{author}{\bibfnamefont{S.}~\bibnamefont{Boseggia}},
  \bibinfo{author}{\bibfnamefont{J.}~\bibnamefont{Akimitsu}},
  \bibinfo{author}{\bibfnamefont{N.~B.} \bibnamefont{Brookes}},
  \bibinfo{author}{\bibfnamefont{M.}~\bibnamefont{Isobe}},
  \bibinfo{author}{\bibfnamefont{M.}~\bibnamefont{Minola}},
  \bibinfo{author}{\bibfnamefont{H.}~\bibnamefont{Okabe}},
  \bibinfo{author}{\bibfnamefont{H.~M.} \bibnamefont{R\o{}nnow}},
  \bibinfo{author}{\bibfnamefont{L.}~\bibnamefont{Simonelli}},
  \bibnamefont{et~al.}, \bibinfo{journal}{Phys. Rev. B}
  \textbf{\bibinfo{volume}{89}}, \bibinfo{pages}{121101}
  (\bibinfo{year}{2014}{\natexlab{b}}),
  \urlprefix\url{https://link.aps.org/doi/10.1103/PhysRevB.89.121101}.

\bibitem[{\citenamefont{Haskel et~al.}(2012)\citenamefont{Haskel, Fabbris,
  Zhernenkov, Kong, Jin, Cao, and van Veenendaal}}]{Haskel2012}
\bibinfo{author}{\bibfnamefont{D.}~\bibnamefont{Haskel}},
  \bibinfo{author}{\bibfnamefont{G.}~\bibnamefont{Fabbris}},
  \bibinfo{author}{\bibfnamefont{M.}~\bibnamefont{Zhernenkov}},
  \bibinfo{author}{\bibfnamefont{P.~P.} \bibnamefont{Kong}},
  \bibinfo{author}{\bibfnamefont{C.~Q.} \bibnamefont{Jin}},
  \bibinfo{author}{\bibfnamefont{G.}~\bibnamefont{Cao}}, \bibnamefont{and}
  \bibinfo{author}{\bibfnamefont{M.}~\bibnamefont{van Veenendaal}},
  \bibinfo{journal}{Phys. Rev. Lett.} \textbf{\bibinfo{volume}{109}},
  \bibinfo{pages}{027204} (\bibinfo{year}{2012}).

\bibitem[{\citenamefont{Zhang et~al.}(2013)\citenamefont{Zhang, Haule, and
  Vanderbilt}}]{Zhang2013}
\bibinfo{author}{\bibfnamefont{H.}~\bibnamefont{Zhang}},
  \bibinfo{author}{\bibfnamefont{K.}~\bibnamefont{Haule}}, \bibnamefont{and}
  \bibinfo{author}{\bibfnamefont{D.}~\bibnamefont{Vanderbilt}},
  \bibinfo{journal}{Phys. Rev. Lett.} \textbf{\bibinfo{volume}{111}},
  \bibinfo{pages}{246402} (\bibinfo{year}{2013}),
  \urlprefix\url{https://link.aps.org/doi/10.1103/PhysRevLett.111.246402}.

\end{thebibliography}

\end{document}